\documentclass{kapproc}

\usepackage{epsf}
\usepackage{graphicx}

\def\be{\begin{equation}}
\def\ee{\end{equation}}
\def\bear{\begin{eqnarray}}
\def\eear{\end{eqnarray}}
\def\bea*{\begin{eqnarray*}}
\def\eea*{\end{eqnarray*}}
\def\bdm{\begin{displaymath}}
\def\edm{\end{displaymath}}

\def\G{\Gamma}
\def\d{\delta}

\def\l{\lambda}

\def\6{\partial}               
\def\7{\llap{/}}               

\def\nn{\nonumber}
\def\<{\langle}
\def\>{\rangle}

\makeatletter
\newbox\slashbox \setbox\slashbox=\hbox{\large$/$}
\def\pslash#1{\setbox\@tempboxa=\hbox{$#1$}
  \@tempdima=0.5\wd\slashbox \advance\@tempdima 0.5\wd\@tempboxa
  \copy\slashbox \kern-\@tempdima \box\@tempboxa}
\def\FMSlash{\protect\pslash}

\normallatexbib

\begin{document}
\articletitle[Quantum chaos in QCD and hadrons]
{
Quantum chaos in QCD and hadrons
}

\author{Harald Markum,\altaffilmark{1} Willibald Plessas,\altaffilmark{2}
Rainer Pullirsch,\altaffilmark{1} Bianka Sengl,\altaffilmark{2} and Robert F. Wagenbrunn\altaffilmark{2}}

\altaffiltext{1}{Atominstitut, Vienna University of Technology,\\
Wiedner Hauptstra\ss e 8-10/141, A--1040 Vienna, Austria}

\altaffiltext{2}{Theoretical Physics, Institute of Physics, 
University of Graz,\\
Universit\"atsplatz 5, A--8010 Graz, Austria}

\begin{abstract}
This article is the written version of a talk delivered at the
Workshop on Nonlinear Dynamics and Fundamental Interactions in Tashkent
and starts with an introduction into quantum chaos
and its relationship to classical chaos. The Bohigas-Giannoni-Schmit
conjecture is formulated and evaluated within random-matrix theory.
In accordance to the title, the presentation is twofold and begins
with research results on quantum chromodynamics and the
quark-gluon plasma.
We conclude with recent research work on the spectroscopy of baryons.
Within the framework of a relativistic constituent quark model we investigate the excitation spectra of the nucleon
and the delta with regard to a possible chaotic behavior for the cases when a hyperfine interaction of either 
Goldstone-boson-exchange or one-gluon-exchange type is added to the confinement interaction.
Agreement with predictions from the experimental hadron spectrum is established.
\end{abstract}

\section{Classical and Quantum Chaos}

\begin{figure}[t]
 \begin{center}
   \hspace{1cm}\includegraphics[width=12cm,height=7cm]{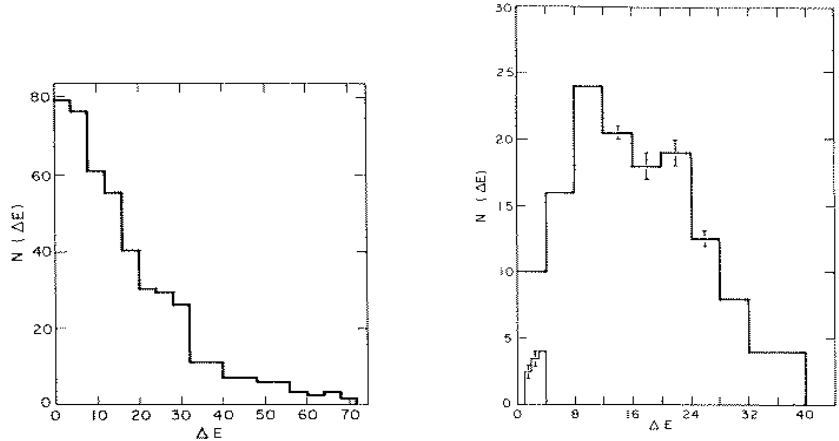}
   \end{center}
   \caption{Nearest-neighbor spacing distributions of eigenvalues
            for a circle (left) and the Bunimovich stadium (right).
            Taken from Ref.~\protect\cite{McDoKauf}.}
   \label{fig2}
 \end{figure}

In order to understand in which manner classical chaos is reflected 
in quantum systems the question has been posed:
Are there differences in the eigenvalue spectra of classically integrable
and non-integrable systems?
Billiards became a preferred playground to study both the classical
and quantum case. With the arrival of computers with increasing
power in the late seventies diagonalization of matrices with
reasonable size became possible. The behavior of the distribution
of the spacings between neighboring eigenvalues turned out to be a
decisive signature. In 1979 McDonald and Kaufman performed a
comparison between the spectra from a classically regular and a
classically chaotic system~\cite{McDoKauf}. As seen in Fig.~\ref{fig2}
they observed a qualitatively different behavior between the nearest-neighbor
spacing distribution of the circle and the stadium. In
the first case the spacings are clearly concentrated around zero
while they show repelling character in the second case.
There were several authors contributing to this discussion and we mention
the papers by Casati, Valz-Gris, and Guarneri~\cite{Casa}, by 
Berry~\cite{Berr}, by Robnik~\cite{Robn}
and by Seligman, Verbaarschot, and Zirnbauer~\cite{SeVeZi}.

\begin{figure}[h]
 \begin{center}
   \hspace{1cm}\includegraphics[width=7cm,height=5cm]{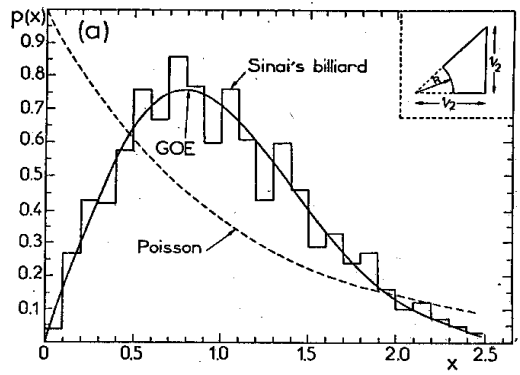}
   \end{center}
   \caption{Nearest-neighbor spacing distributions of eigenvalues for
            the Sinai billiard with the Wigner surmise compared to the
            Poisson distribution. The histogram comprises about 1000
            consecutive eigenvalues. Taken from Ref.~\protect\cite{Bohi84}.}
   \label{fig3}
 \end{figure}

Very accurate results were obtained for the classically chaotic Sinai
billiard by Bohigas, Giannoni, and Schmit (see Fig.~\ref{fig3}) which
led them to the important conclusion~\cite{Bohi84}:
Spectra of time-reversal invariant systems whose classical analogues are
K systems show the same fluctuation properties as predicted by the Gaussian
orthogonal ensemble (GOE) of random-matrix theory (RMT).  K systems are
most strongly mixing classical systems with a positive Kolmogorov entropy.
The conjecture turned out valid also for less chaotic (ergodic) systems
without time-reversal invariance leading to the Gaussian unitary ensemble
(GUE).

\section{Random Matrix Theory}
\label{sec3}

In lack of analytical or numerical methods to obtain the spectra of complicated
Hamiltonians, Wigner and Dyson analyzed ensembles of random matrices and
were able to derive mathematical expressions.
A Gaussian random matrix ensemble consists of square matrices with their
matrix elements drawn from a Gaussian distribution
\begin{equation}
p(x) = \frac{1}{\sqrt{2\pi}\sigma} \; \exp\left(-\frac{x^2}{2\sigma^2}\right) \ .
\end{equation}
One distinguishes between three different types depending on space-time
symmetry classified by the Dyson parameter $\beta_D = 1, 2, 4$~\cite{Guhr}.
The Gaussian orthogonal ensemble (GOE, $\beta_D = 1$) holds for
time-reversal invariance and rotational symmetry of the Hamiltonian
\bear
H_{mn} = H_{nm} = H^{\ast}_{nm} \ .
\eear
When time-reversal invariance is violated and
\bear
H_{mn}=[H^{\dagger}]_{mn} \ ,
\eear
one obtains the Gaussian unitary ensemble (GUE, $\beta_D = 2$).
The Gaussian symplectic ensemble (GSE, $\beta_D = 4$) is in 
correspondence with time-reversal invariance but broken rotational
symmetry of the Hamiltonian
\bear
H^{(0)}_{nm}\mbox{1\hspace*{-2mm}I}_2 - i \sum\limits_{\gamma=1}^3 
H^{(\gamma)}_{nm} \sigma_{\gamma} \ ,
\eear
with $H^{(0)}$ real and symmetric and $H^{(\gamma)}$ real and antisymmetric.

The functional form of the distribution $P(s)$ of the neighbor
spacings $s$ between consecutive eigenvalues for the Gaussian
ensembles can be approximated by 
\bear
\label{wignersurmise}
P_{\beta_D}(s) = a_{\beta_D} \, s^{\beta_D} \exp\left(-b_{\beta_D} \,
s^2 \,\right) \ ,
\eear
which is known as the Wigner surmise and reads for example in the case
$\beta_D =2$ (GUE)
\bear
P(s) = \frac{32}{\pi^2} \, s^2 \exp\left(-\frac{4}{\pi} \, s^2 \right) \ .
\eear
If the eigenvalues of a system are completely uncorrelated one ends up
with a Poisson distribution for their neighbor spacings
\bear
P(s) = \exp{(-s)} \ .
\eear
An interpolating function between the Poisson and the Wigner distribution
is given by the Brody distribution~\cite{Brod} reading for the GOE case
\bear
\label{brodydist}
P(s,\omega) = \alpha \, (\omega +1)  s^\omega 
\exp\left(- \alpha \, s^{\omega + 1}\right) \ ,
\quad \quad \alpha = \Gamma^{\omega+1} \,
\left(\frac{\omega + 2}{\omega + 1}\right) \ ,
\eear
with $0 \leq \omega \leq 1$.

Remarkably, the Wigner distribution could
be observed in a number of systems by physical experiments and computer simulations
evading the whole quantum world
from atomic nuclei to the hydrogen atom in a magnetic field to the metal-insulator
transition~\cite{Guhr}.
In this contribution we address the situation in QCD and in hadrons.

\section{Quantum Chromodynamics}

The Lagrangian ${\cal L}^{\mbox{\scriptsize QCD}}$ of quantum chromodynamics
(QCD) consists of a gluonic part
${\cal L}_{\mbox{\scriptsize G}}^{\mbox{\scriptsize QCD}}$ and a part
${\cal L}_{\mbox{\scriptsize F}}^{\mbox{\scriptsize QCD}}$ from the quarks
\bear
{\cal L}^{\mbox{\scriptsize QCD}} & = &
{\cal L}_{\mbox{\scriptsize G}}^{\mbox{\scriptsize QCD}}+
{\cal L}_{\mbox{\scriptsize F}}^{\mbox{\scriptsize QCD}} \nn \\
& = &
- \frac{1}{4} F_{\mu\nu}^{a}(x)F_{a}^{\mu\nu}(x) +
\sum\limits_{f=1}^{N_{f}} \bar\psi_{f}(x)(i D\!\7 - m_{f}) \psi_{f}(x) \ ,
\eear
with the Dirac spinor $\psi_{f}$, the quark mass $m_{f}$, the number of
flavors $N_{f}$, and the generalized field strength tensor
\bear
F_{a}^{\mu\nu}(x) = \6^{\mu}A_{a}^{\nu}(x)-\6^{\nu}A_{a}^{\mu}(x)
- g f_{abc}A_{b}^{\mu}(x)A_{c}^{\nu}(x)        \ ,
\eear
where the gauge field $A_{a}^{\mu}$ with the SU(3) indices
$a,b,c=1, \dots ,8$, the coupling constant $g$ and the structure
constants $f_{abc}$ of SU(3) enter.
The main object of study is the eigenvalue spectrum of the Dirac
operator of QCD in 4 dimensions
\bear
\FMSlash{D} = \FMSlash{\partial} + i g \FMSlash{A}^a
\frac{\lambda^a}{2} = \gamma_{\mu} \partial_{\mu} + i g \gamma_{\mu} A^a_{\mu}
\frac{\lambda^a}{2} \ ,
\eear
with the $\l_{a}$ the generators of the SU(3) color-group (Gell-Mann matrices).
Discretizing the Dirac operator on a lattice in Euclidean space-time
and applying the Kogut-Susskind (staggered) prescription, leads to the matrix
\bear
(M_{\mbox{\scriptsize KS}})_{xx'}^{aa'} =
\frac{1}{2a} \ \sum_{\mu} \left[
\d_{x+\hat\mu,x'} \ \G_{x\mu} \ U_{x\mu}^{aa'} - \d_{x,x'+\hat\mu}
\ \G_{x'\mu} \ U^{\dagger \ aa'}_{x'\mu} \right] \ ,
\eear
where 
\bear
U_{x \mu} =  \exp \left(igA_{\mu}^{a}(x)\frac{\l^{a}}{2} \ \right) 
\eear
are the gauge field variables on the lattice and $\Gamma_{x\mu}$ a
representation of the $\gamma_{\mu}$-matrices.

In random matrix theory (RMT),
one has to distinguish several universality classes which are
determined by the symmetries of the system.  For the case of the QCD
Dirac operator, this classification was done in
Ref.~\cite{Verb94}.  Depending on the number of colors and the
representation of the quarks, the Dirac operator is described by one
of the three chiral ensembles of RMT.  As far as the fluctuation
properties in the bulk of the spectrum are concerned, the predictions
of the chiral ensembles are identical to those of the ordinary
ensembles in Sect.~\ref{sec3}~\cite{Fox64}.
In Ref.~\cite{Hala95}, the Dirac matrix was
studied for color-SU(2) using both Kogut-Susskind and Wilson fermions which
correspond to the chiral symplectic (chSE) and orthogonal (chOE) ensemble,
respectively. Here~\cite{Pull98}, we additionally study SU(3) with
Kogut-Susskind fermions which corresponds to the chiral unitary ensemble (chUE).
The RMT result for the nearest-neighbor spacing distribution can be
expressed in terms of so-called prolate spheroidal functions, see
Ref.~\cite{Meht91}.  A very good approximation to $P(s)$ is
provided by the Wigner surmise for the unitary ensemble,
\begin{equation} \label{wigner}
  P_{\rm W}(s)=\frac{32}{\pi^2}s^2e^{-4s^2/\pi} \:.
\end{equation}

We generated gauge field configurations using the standard Wilson
plaquette action for SU(3) with and without dynamical fermions in the
Kogut-Susskind prescription. We have worked on a $6^3\times 4$ lattice
with various values of the inverse gauge coupling $\beta=6/g^2$ both
in the confinement and deconfinement phase.  We typically produced 10
independent equilibrium configurations for each $\beta$.  Because of
the spectral ergodicity property of RMT one can replace ensemble
averages by spectral averages if one is only interested in bulk
properties.

The Dirac operator, $\FMSlash{D}=\FMSlash{\partial}+ig\FMSlash{A}$, is
anti-Hermitian so that the eigenvalues $\lambda_n$ of $i\FMSlash{D}$
are real.  Because of $\{\FMSlash{D},\gamma_5\}=0$ the non-zero
$\lambda_n$ occur in pairs of opposite sign.  All spectra were checked
against the analytical sum rules $\sum_{n} \lambda_n = 0$ and
$\sum_{\lambda_n>0} \lambda_n^2 = 3V$, where V is the lattice volume.
To construct the nearest-neighbor spacing distribution from the
eigenvalues, one first has to ``unfold'' the spectra~\cite{Bohi84a}.

\begin{figure}[hp]
\begin{center}
\begin{tabular}{ccccc}
  & {\large Confinement $\beta=5.2$}  & \hspace*{6.5mm}   & &
  {\large Deconfinement $\beta=5.4$} \\
  \vspace*{0mm}
  & {\large $ma=0.05$} &&& {\large $ma=0.05$} \\[2mm]
  \multicolumn{2}{c}{\epsfxsize=5cm\epsffile{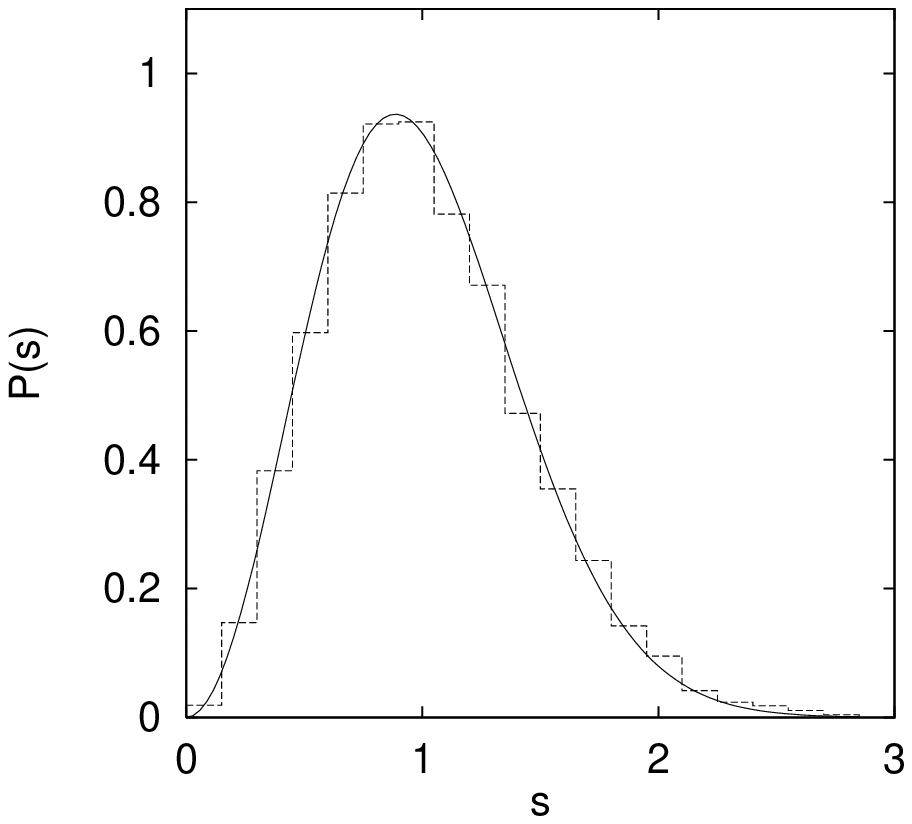}} &&
  \multicolumn{2}{c}{\epsfxsize=5cm\epsffile{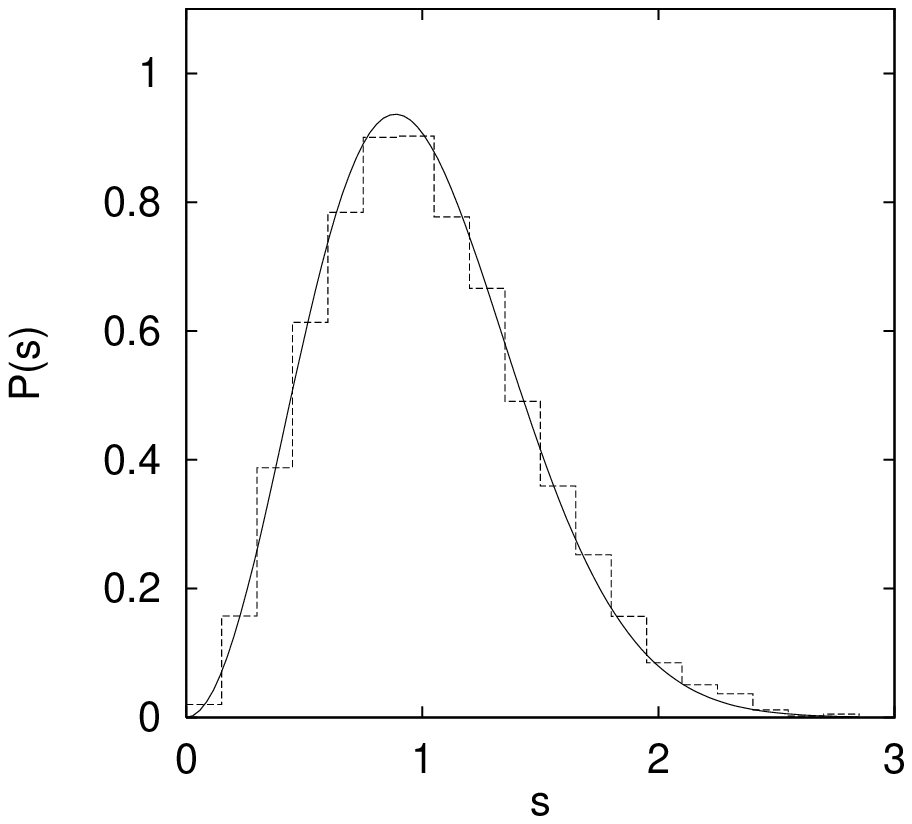}}
\end{tabular}  
\end{center}
\caption{Nearest-neighbor spacing distribution $P(s)$ for the Dirac
  operator on a $6^3 
  \times 4$ lattice in full QCD (histograms) compared with the
  random matrix result (solid lines). There are no changes in $P(s)$
  across the deconfinement phase transition.}
\vspace*{8mm}
\label{fintemp}
  \centerline{
  \includegraphics[width=5cm]{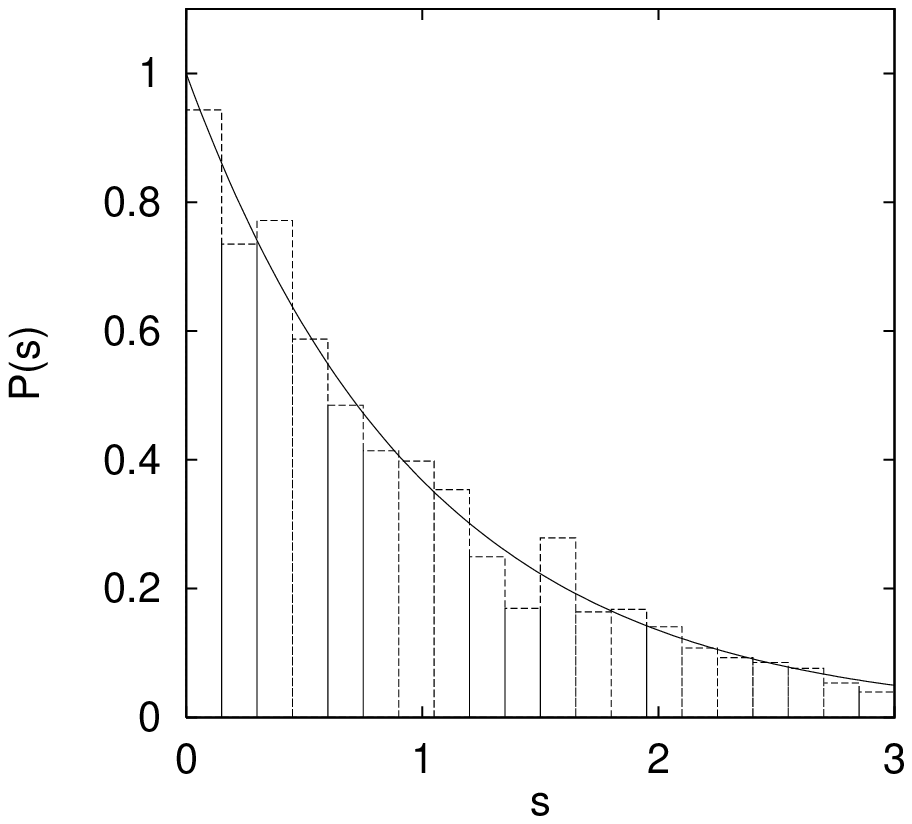}
  }
  \caption{Nearest-neighbor spacing distribution $P(s)$ for the free
            Dirac operator on a $53\times 47\times 43\times 41$ lattice
            compared with a Poisson distribution, $e^{-s}$.}
  \label{free}
\end{figure}

Figure~\ref{fintemp} compares $P(s)$ of full QCD with $N_f = 3$
flavors and quark mass $ma=0.05$ to the RMT result.  In the confinement
as well as in the deconfinement phase we observe agreement with RMT up
to very high $\beta$ (not shown).  The observation that $P(s)$ is not
influenced by the presence of dynamical quarks is
expected from the results of Ref.~\cite{Fox64}, which
apply to the case of massless quarks. Our
results, and those of Ref.~\cite{Hala95}, indicate that massive
dynamical quarks do not affect $P(s)$ either.

No signs for a transition to Poisson regularity are found. The
deconfinement phase transition does not seem to coincide with a
transition in the spacing distribution. For very large values of
$\beta$ far into the deconfinement region, the eigenvalues
start to approach the degenerate eigenvalues of the free theory, given
by $\lambda^2=\sum_{\mu=1}^4 \sin^2(2\pi n_\mu/L_\mu)/a^2$, where $a$
is the lattice constant, $L_{\mu}$ is the number of lattice sites in
the $\mu$-direction, and $n_\mu=0,\ldots,L_\mu-1$.  In this case, the
nearest-neighbor spacing distribution is neither Wigner nor Poisson.
It is possible to lift the degeneracies of the free
eigenvalues using an asymmetric lattice where $L_x$, $L_y$, etc. are
relative primes and, for large lattices, the distribution 
is then Poisson, $P_{\rm P}(s)=e^{-s}$, see Fig.~\ref{free}.

\begin{figure*}[t]
  \centerline{\includegraphics[width=5cm]{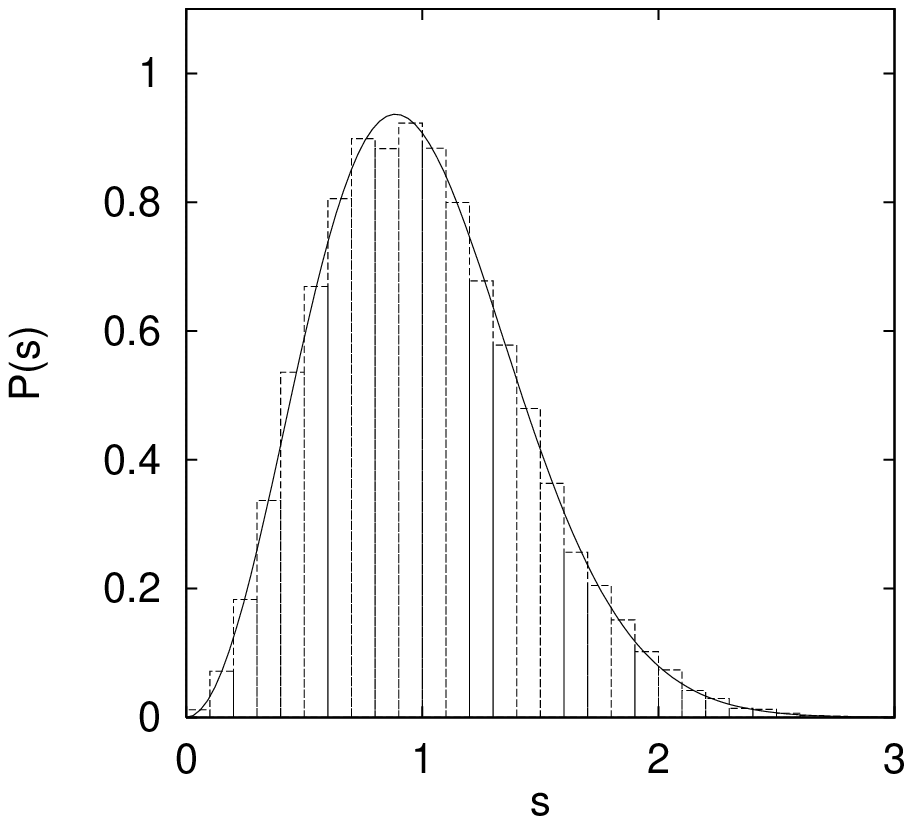}\hspace*{10mm}
    \includegraphics[width=5cm]{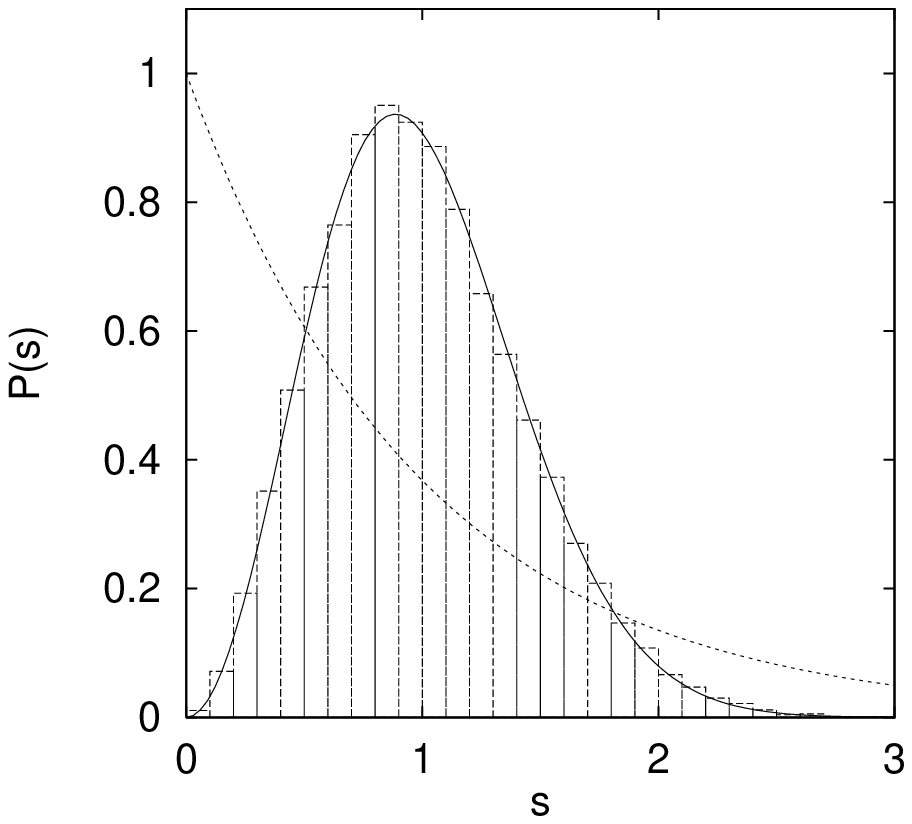}}
  \caption{Nearest-neighbor spacing distribution $P(s)$ for U(1) gauge
    theory on an $8^3\times 6$ lattice in the confined phase (left)
    and in the Coulomb phase (right). The theoretical curves are the chUE
    result, Eq.~(\ref{wigner}), and the Poisson distribution, $P_{\rm
      P}(s)=\exp(-s)$.}
  \label{f02}
\end{figure*}

We have also investigated the staggered Dirac spectrum of 4d U(1)
gauge theory which corresponds to the chUE of RMT but had
not been studied before in this context.  At $\beta_c \approx 1.01$
U(1) gauge theory undergoes a phase transition between a confinement
phase with mass gap and monopole excitations for $\beta < \beta_c$ and
the Coulomb phase which exhibits a massless photon for $\beta >
\beta_c$. As for SU(2) and SU(3) gauge groups, we
expect the confined phase to be described by RMT, whereas free
fermions are known to yield the Poisson distribution (see
Fig.~\ref{free}). The question arose whether the Coulomb phase would be
described by RMT or by the Poisson distribution~\cite{BeMaPu99}. The
nearest-neighbor spacing distributions for an $8^3\times 6$ lattice at
$\beta=0.9$ (confined phase) and at $\beta=1.1$ (Coulomb phase),
averaged over 20 independent configurations, are depicted in
Fig.~\ref{f02}. Both are consistent with the chUE of RMT.

\section{Hadrons}

\begin{figure}[h]
\begin{center}
\begin{tabular}{lcr}
\epsfxsize=3.5cm\epsffile{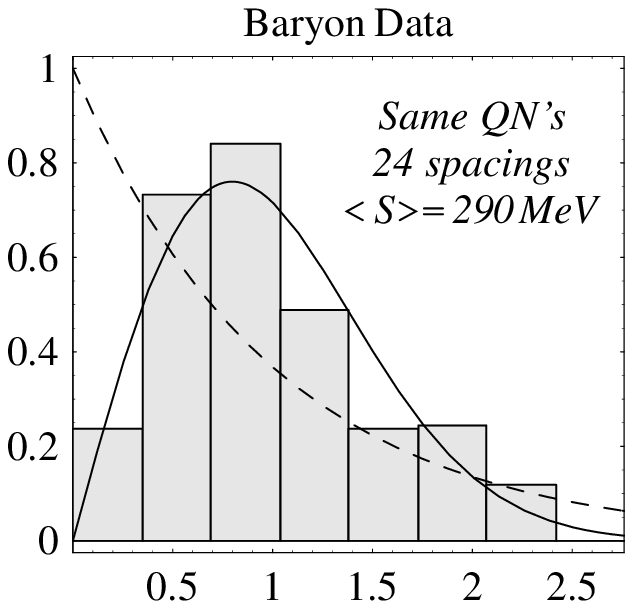} &
\epsfxsize=3.5cm\epsffile{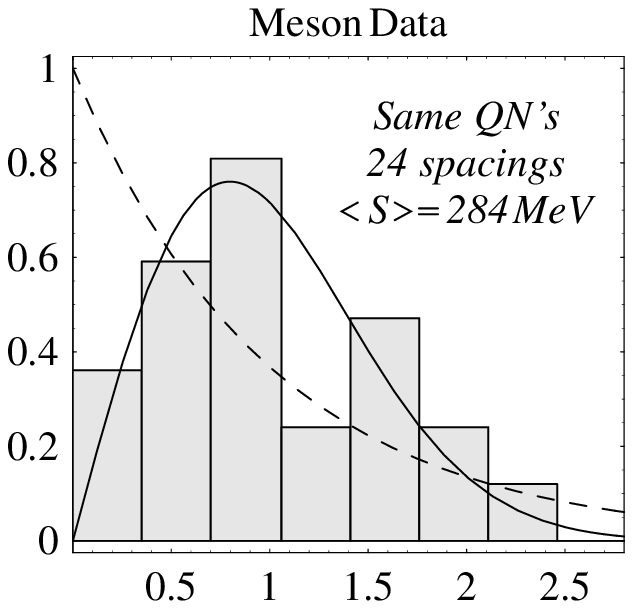} &
\epsfxsize=3.5cm\epsffile{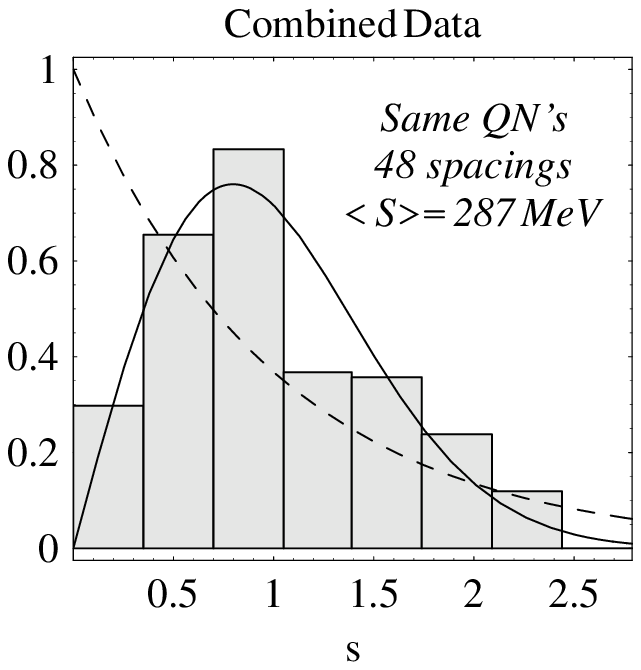} \\
\end{tabular}
\end{center}
\caption{Histograms of the nearest-neighbor mass spacing
distribution for hadron states with same quantum numbers.
Curves represent the Poisson (dashed) and Wigner (solid)
distributions. Taken from Ref.~\protect\cite{Pasc}.}
\label{pofs}
\end{figure}

Taking the experimentally measured mass spectrum of hadrons
up to 2.5 GeV from the Particle Data Group, Pascalutsa~\cite{Pasc}
could show that the hadron level-spacing
distribution is remarkably well described by the
Wigner surmise for $\beta=1$ (see Fig.~\ref{pofs}). This indicates that
the fluctuation properties of the hadron spectrum
fall into the GOE universality class, and
hence hadrons exhibit the {\it quantum chaos} phenomenon.
One then should be able to describe
the statistical properties of hadron spectra
using RMT with random Hamiltonians from GOE
that are characterized by good time-reversal
and rotational symmetry.

\begin{figure}
\begin{center}
$\includegraphics[height=6cm]{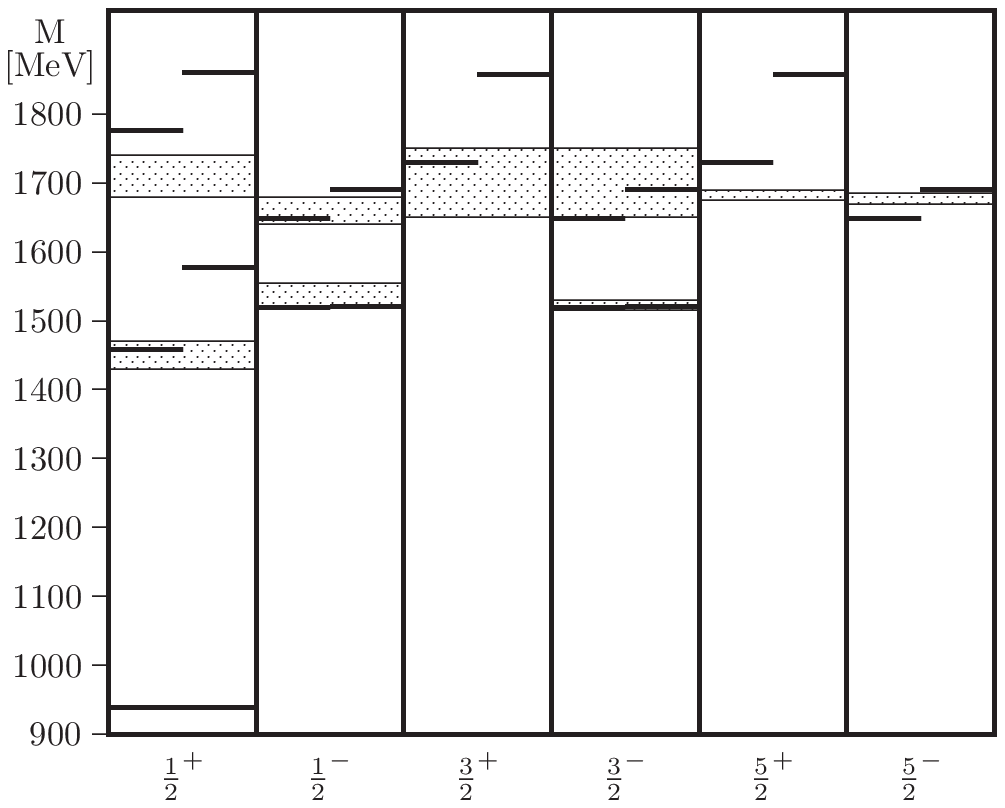}\hfill\includegraphics[height=6cm]{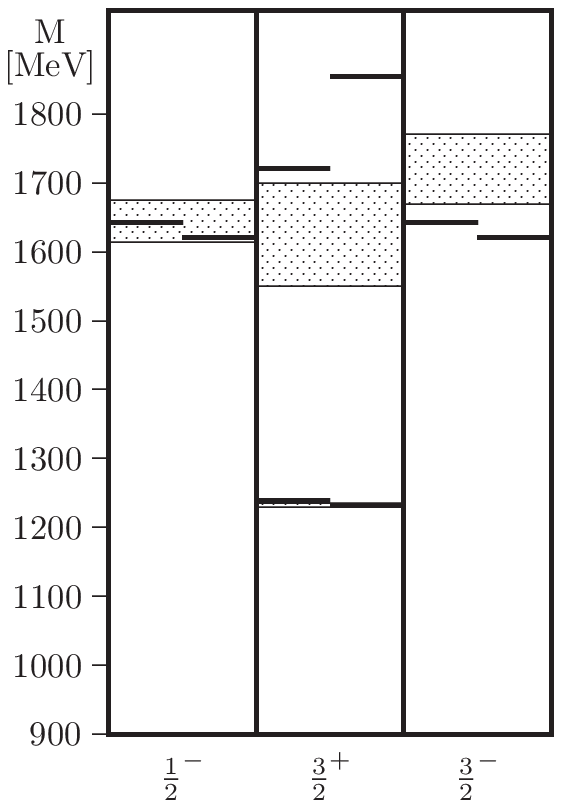}$
\end{center}
\caption{Low lying nucleon (left plot) and delta (right plot)
states with total spin and parity $J^\pi$. The left and right bars are the
theoretical energies predicted from the GBE and OGE models as described in the
text, respectively.
The shaded boxes represent the experimental energies with their uncertainties
\protect\cite{pdg}.}
\label{spectrum}
\end{figure}

In order to test this experimental finding we are comparing with
the eigenvalues of a Hamiltonian for a realistic quark model,
namely the Goldstone-boson-exchange (GBE) constituent quark model
\cite{Glozman98}. It includes 
the kinetic energy in relativistic form
\begin{equation}
\label{kin_sr}
H_0=\sum\limits_{i=1}^3 \sqrt{\vec{p}_i^{\;2}+m_i^2},
\end{equation}
with $m_i$ the masses and $\vec{p}_i$ the 3-momenta of the constituent quarks.
The
interaction between two constituent quarks $i,j$
\begin{equation}
\label{vgbe}
V(ij)=V_{\rm conf}(ij)+V_{\chi}(ij)
\end{equation}
is given by a confinement potential in linear form
\begin{equation}
\label{vconf}
V_{\rm conf}(ij)=V_0+C r_{ij}
\end{equation}
and a hyperfine interaction consisting of only the spin-spin part of the
pseudoscalar-meson-exchange potentials
\begin{equation}
\label{vchi}
\begin{array}{l}
V_\chi(ij)  =
\left[\sum_{F=1}^3 V_{\pi}(r_{ij}) \lambda_i^F \lambda_j^F\right.
+\sum_{F=4}^7 V_K(r_{ij}) \lambda_i^F \lambda_j^F\\[2ex]
\hspace{4.5cm}\left.+V_{\eta}(r_{ij}) \lambda_i^8
\lambda_j^8 + \frac{2}{3}V_{\eta'}(r_{ij})\right]
\vec\sigma_i\cdot\vec\sigma_j.
\end{array}
\end{equation}
Here $r_{ij}$ is the distance between the quarks, $\vec\sigma_i$
are the Pauli spin matrices and $\lambda_i$
the Gell-Mann flavor matrices of the individual quarks.
This kind of interaction is motivated by the spontaneous breaking of
chiral symmetry. As a consequence constituent quarks and Goldstone bosons
should be the appropriate effective degrees of freedom at low energies.
Baryons are then assumed to be bound states of three confined constituent
quarks with a hyperfine interaction relying on
the exchange of the Goldstone bosons.
Due to the specific flavor dependence in Eq. (\ref{vchi}) a reasonable
agreement between the spectra of the low lying light and strange baryon
states calculated from the model and the experimental spectra could
be achieved. In particular the ordering of the excited states
with respect to their parities comes out correctly as is demonstrated
in Fig.~\ref{spectrum}.
It is interesting to notice that both the experiment and the numerical
treatment have their problems to resolve the higher excited states.

In order to investigate the influence of the hyperfine interaction we also
analyze the nearest-neighbor spacings obtained with the
confinement potential without the hyperfine interaction $V_{\chi}$
and with a model consisting of a different kind of hyperfine
interaction which is based on one-gluon exchange (OGE). This was traditionally
used in constituent quark models and has a flavor independent
spin-spin potential. Therefore it has principal problems in reproducing
the phenomenological ordering of the low lying excited nucleon states.
Nevertheless, for comparison we consider here a simple version of such a model,
i.e., a reparametrization of the Bhaduri, Cohler, and Nogami model consisting
of a potential of the form
\begin{equation}
\label{bhaduri}
V(ij)=V_0+C r_{ij}-\frac{2 b}{3 r_{ij}} +\frac{\alpha_s}{9 m_im_j}\Lambda^2
       \frac{e^{-\Lambda r_{ij}}}{r_{ij}}\vec{\sigma}_{i}\vec{\sigma}_{j},
\end{equation}
and also a relativistic kinetic energy term in its Hamiltonian \cite{Theussl}.
The spectra of the low lying nucleon and delta states calculated with this
model are inserted in Fig.~\ref{spectrum}.

\begin{figure}[hp]
\begin{center}
\begin{tabular}{lr}
\includegraphics[width=5cm,height=5cm,angle=-90]{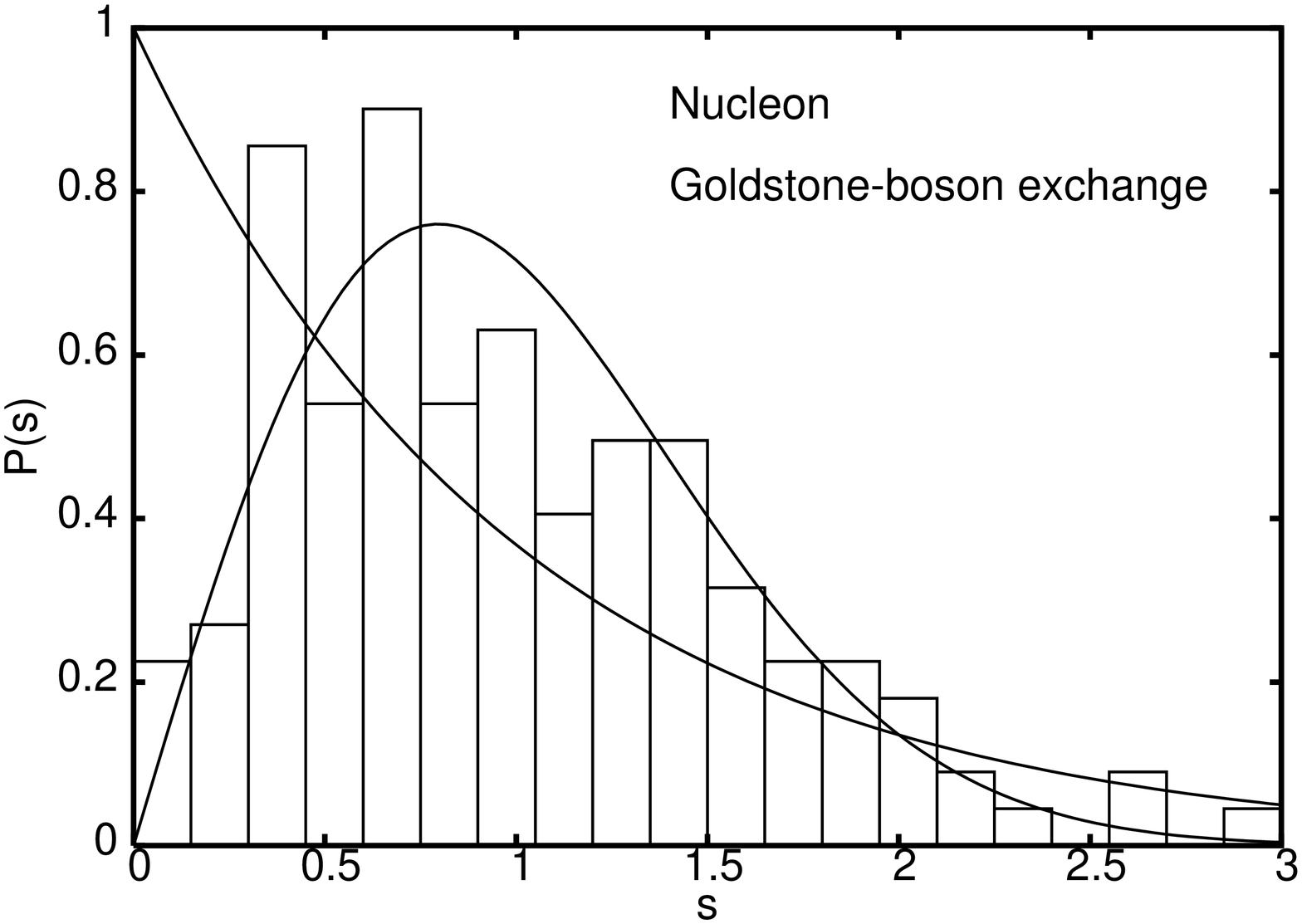} &
\includegraphics[width=5cm,height=5cm,angle=-90]{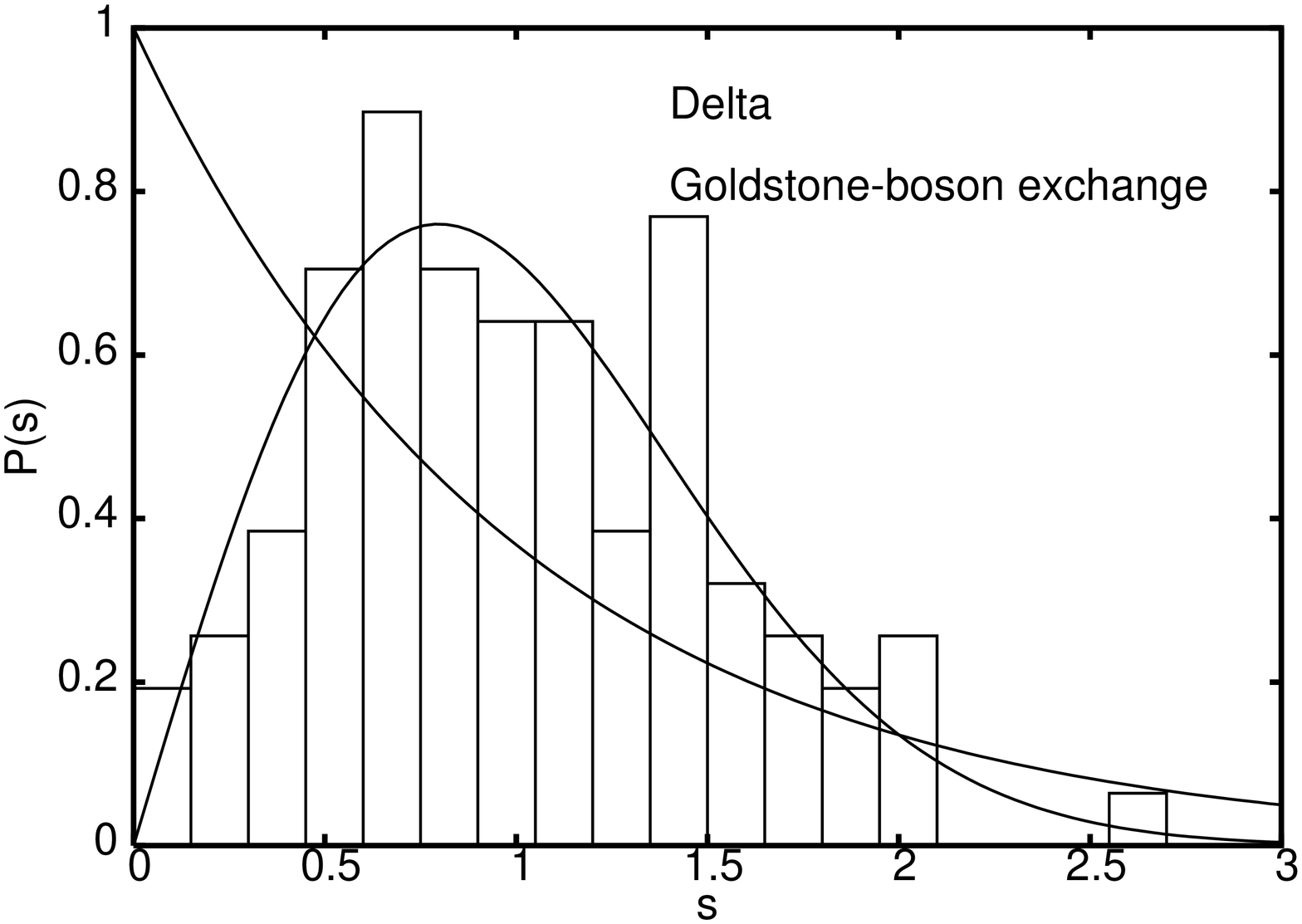} \\
\includegraphics[width=5cm,height=5cm,angle=-90]{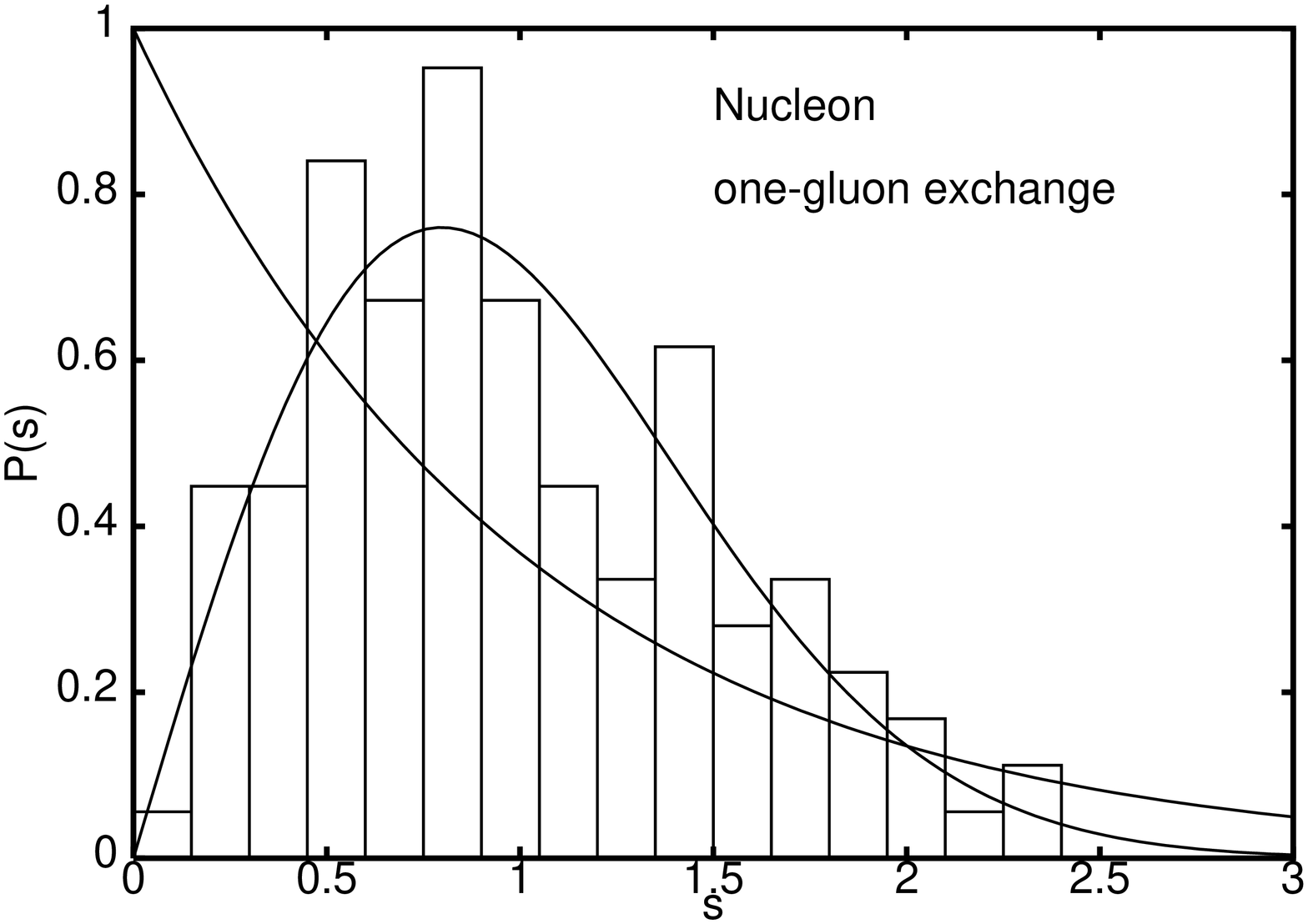} &
\includegraphics[width=5cm,height=5cm,angle=-90]{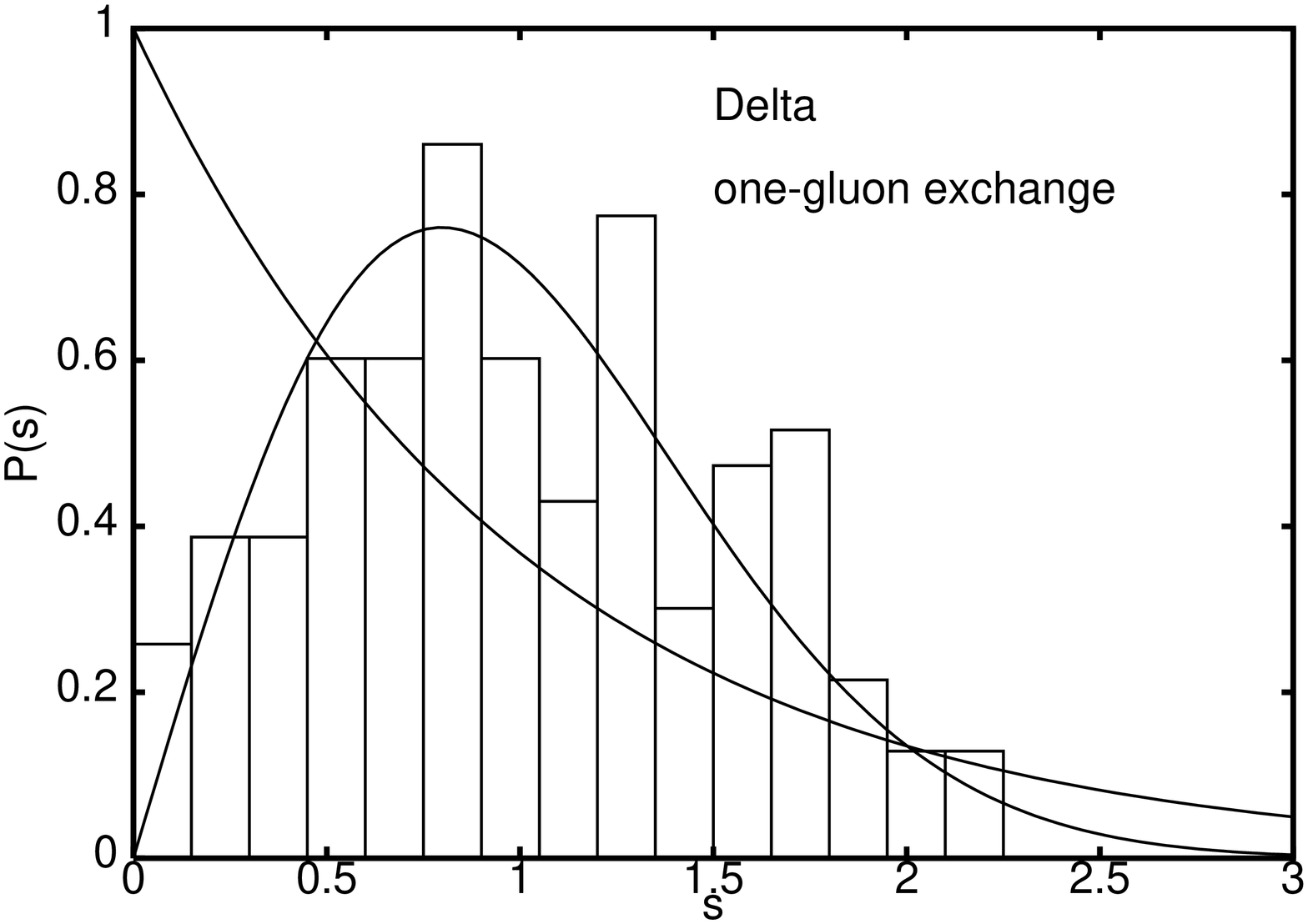} \\
\includegraphics[width=5cm,height=5cm,angle=-90]{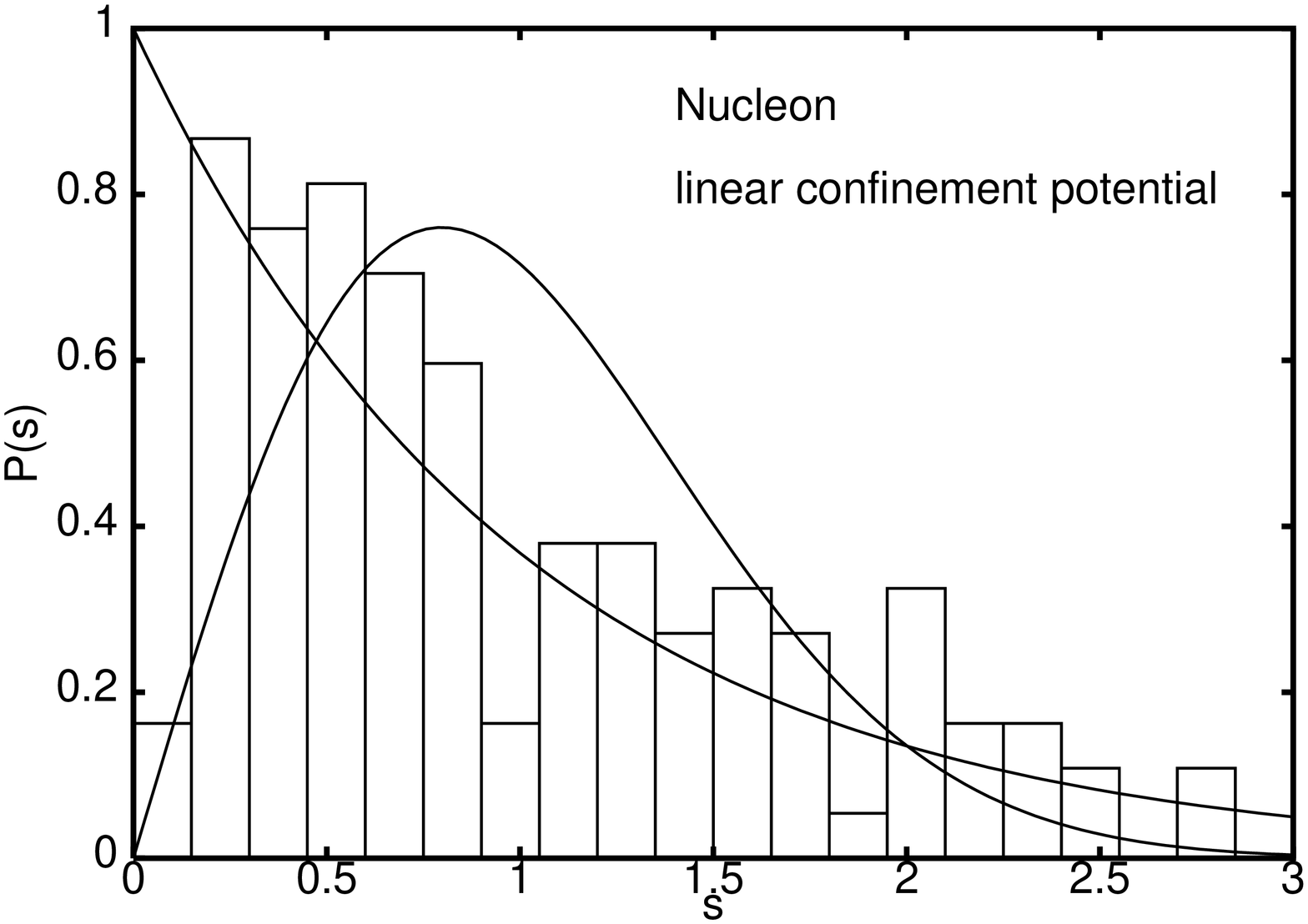} &
\includegraphics[width=5cm,height=5cm,angle=-90]{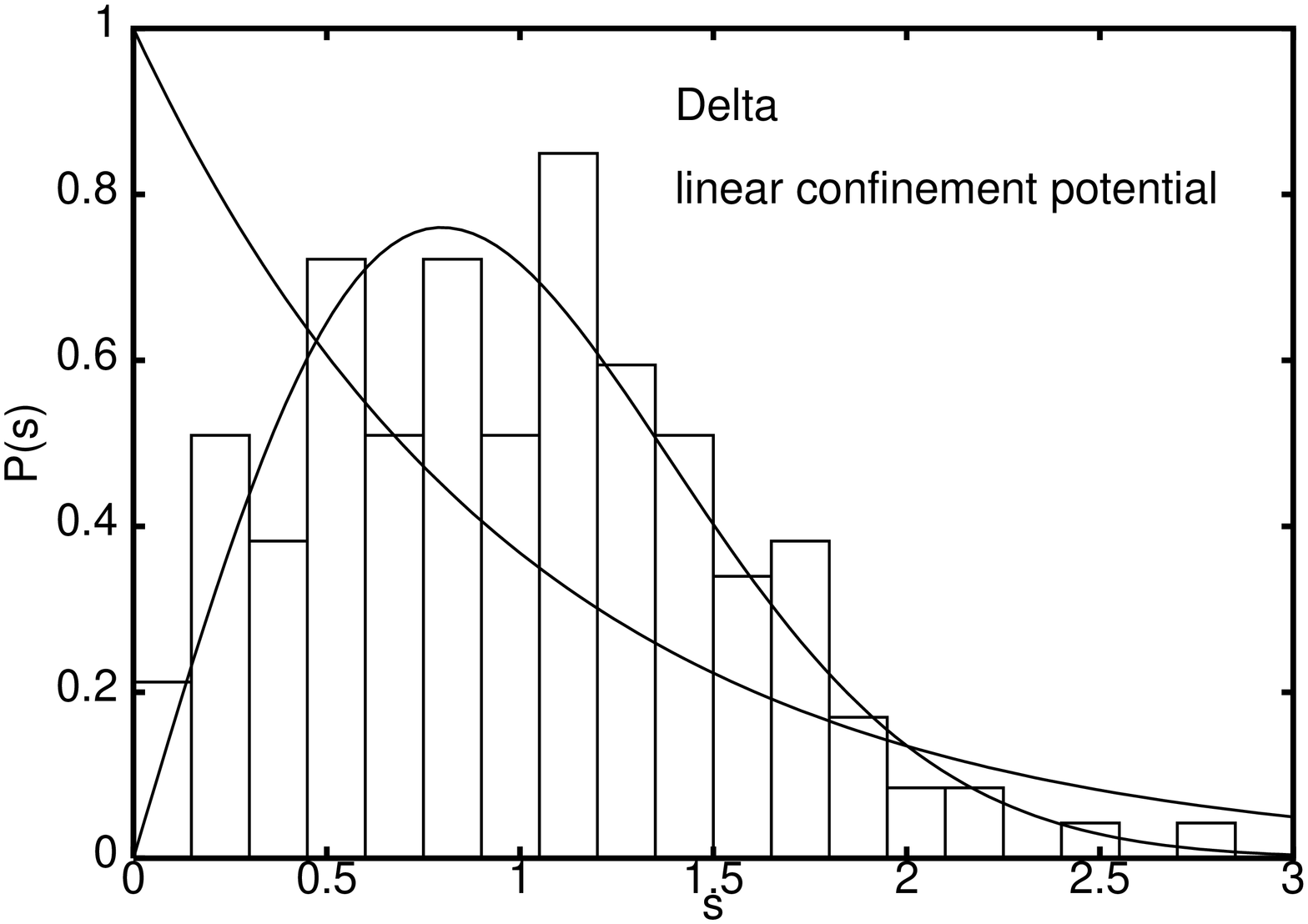} \\
\end{tabular}
\end{center}
\vspace*{-3mm}
  \caption{Histograms of the nearest-neighbor spacing distribution 
           for the nucleon (left plots) and the delta (right plots).
           The data is for Goldstone-boson exchange and for one-gluon exchange
           compared to a pure linear confinement potential of the same strength.
           Curves represent the Poisson and the GOE-Wigner distributions.
    }
  \vspace*{-3mm}
  \label{pshadron}
\end{figure}
In Fig.~\ref{pshadron} we present our theoretical results of the
nearest-neighbor spacing distribution for the nucleon and the delta.
Both the hyperfine interaction of either Goldstone-boson exchange
and one-gluon exchange type yield spacing distributions
corresponding to the GOE.
One observes a preference for the GOE from the linear rise at the
origin while the other ensembles are quadratic or quartic 
(cf.~Eq.~\ref{wignersurmise}).
It turns out that the linear confinement potential alone without
reproducing the spectra yields eigenvalues with reduced correlations
between their neighbors and thus leading towards a Poisson distribution,
as seen clearly from the nucleon in Fig.~\ref{pshadron}.

\section{Conclusion}

We have outlined the universal applicability of random-matrix theory
and have presented own studies of quantum chromodynamics and hadrons.
Concerning QCD, we were able to
demonstrate that the nearest-neighbor spacing distribution $P(s)$ of the
eigenvalues of the Dirac operator agrees perfectly with the RMT prediction
both in the confinement and quark-gluon plasma-phase. This means that
QCD is governed by quantum chaos in both phases. We could show that the
eigenvalues of the free Dirac theory yield a Poisson distribution related
to regular behavior. Our investigations tell us that the critical point of
the spontaneous breaking of chiral symmetry does not coincide with a
chaos-to-order transition.
Concerning quarks building hadrons, we employed a relativistic quark
potential model allowing for meson exchange or one-gluon exchange.
Computing the spectrum of the nucleon and 
delta baryon indicates a spacing distribution $P(s)$ favoring the GOE
of RMT. A linear confinement potential alone without reproducing the 
level ordering is not enough to obtain the correct fluctuations 
between the eigenvalues. Our results are in agreement with an analysis
of the experimental mass spectrum of hadrons from the Particle Data Tables.
Invoking the Bohigas-Giannoni-Schmit conjecture, we conclude that not
only the quarks but also the hadrons show evidence of quantum
chaos.


\end{document}